\documentclass{aa}

\pdfoutput=1

\usepackage{graphicx}
\usepackage{txfonts}
\usepackage[colorlinks=true,urlcolor=blue,citecolor=blue]{hyperref}

\newcommand{\age}{\ensuremath{200^{+100}_{-50}}}
\newcommand{\teff}{\ensuremath{8980^{+90}_{-130}}}
\newcommand{\vsini}{\ensuremath{114 \pm 3}}
\newcommand{\logg}{\ensuremath{4.31 \pm 0.02}}
\newcommand{\metal}{\ensuremath{-0.02 \pm 0.07}}
\newcommand{\mstar}{\ensuremath{1.89^{+0.06}_{-0.05}}}
\newcommand{\rstar}{\ensuremath{1.60 \pm 0.06}}
\newcommand{\sdens}{\ensuremath{0.66 \pm 0.01}}

\newcommand{\mplan}{\ensuremath{< 17}}
\newcommand{\rplan}{\ensuremath{1.83 \pm 0.07}}
\newcommand{\teq}{\ensuremath{2260 \pm 50}}

\newcommand{\epoch}{\ensuremath{2457909.5906^{+0.0003}_{-0.0002}}}
\newcommand{\period}{\ensuremath{3.474119^{+0.000005}_{-0.000006}}}
\newcommand{\axis}{\ensuremath{0.057 \pm 0.006}}
\newcommand{\inc}{\ensuremath{86.4^{+0.5}_{-0.4}}}
\newcommand{\ecc}{\ensuremath{0~\rm{(fixed)}}}
\newcommand{\obl}{\ensuremath{0.6 \pm 4}}

\usepackage{etoolbox}
\makeatletter
\patchcmd\@combinedblfloats{\box\@outputbox}{\unvbox\@outputbox}{}{%
   \errmessage{\noexpand\@combinedblfloats could not be patched}%
}%
 \makeatother

\begin{document}

\title{MASCARA-2\,b}
\subtitle{A hot Jupiter transiting the $m_V=7.6$ A-star HD\,185603.}
\author{G.J.J. Talens \inst{1}
\and A. B. Justesen \inst{2}
\and S. Albrecht \inst{2}
\and J. McCormac \inst{5}
\and V. Van Eylen \inst{1}
\and G.P.P.L. Otten \inst{1}
\and F. Murgas \inst{3,4}
\and E. Palle \inst{3,4}
\and D. Pollacco \inst{5}
\and R. Stuik \inst{1}
\and J.F.P. Spronck \inst{1}
\and A.-L. Lesage \inst{1}
\and F. Grundahl \inst{2}
\and M. Fredslund Andersen \inst{2}
\and V. Antoci \inst{2}
\and \\I.A.G. Snellen \inst{1}
}
\institute{Leiden Observatory, Leiden University, Postbus 9513, 2300 RA, Leiden, The Netherlands\\
\email{talens@strw.leidenuniv.nl}
\and Stellar Astrophysics Centre (SAC), Department of Physics and Astronomy, Aarhus University, Ny Munkegade 120, DK-8000
\and Instituto de Astrof\'{i}sica de Canarias (IAC), V\'ia L\'actea s/n, 38205, La Laguna, Tenerife, Spain
\and Departamento de Astrof\'{i}sica, Universidad de La Laguna, 38205, La Laguna, Tenerife, Spain
\and Department of Physics, University of Warwick, Coventry CV4 AL, UK
}

\abstract{In this paper we present MASCARA-2\,b, a hot Jupiter transiting the $m_V=7.6$ A2 star HD 185603. Since early 2015, MASCARA has taken more than 1.6 million flux measurements of the star, corresponding to a total of almost 3000 hours of observations, revealing a periodic dimming in the flux with a depth of $1.3\%$. Photometric follow-up observations were performed with the NITES and IAC80 telescopes and spectroscopic measurements were obtained with the Hertzsprung SONG telescope. We find MASCARA-2\,b orbits HD\,185603 with a period of $\period~\rm{days}$ at a distance of $\axis~\rm{AU}$, has a radius of $\rplan~\rm{R}_{\rm{J}}$ and place a $99\%$ upper limit on the mass of $\mplan~\rm{M}_{\rm{J}}$. HD\,185603 is a rapidly rotating early-type star with an effective temperature of \teff~K and a mass and radius of \mstar~$M_\odot$, \rstar~$R_\odot$, respectively. Contrary to most other hot Jupiters transiting early-type stars, the projected planet orbital axis and stellar spin axis are found to be aligned with $\lambda=\obl\degr$. The brightness of the host star and the high equilibrium temperature, $\teq~\rm{K}$, of MASCARA-2\,b make it a suitable target for atmospheric studies from the ground and space. Of particular interest is the detection of TiO, which has recently been detected in the similarly hot planets WASP-33\,b and WASP-19\,b.}

\maketitle

\section{Introduction}
\label{sec:introduction}

Since the first discoveries of extrasolar planets in the mid-nineties the number of known exoplanets has been growing rapidly, first through radial velocity surveys \citep{Vogt2000, Valenti2005} and later through the first succesfull gound-based transit surveys such as HAT \citep{Bakos2004} and SuperWASP \citep{Pollacco2006}. These early surveys uncovered a population of gas giant planets at small orbital distances, the so-called hot Jupiters, of which a few hundred are known today. More recently the space-based transit surveys, CoRoT \citep{Barge2008} and in particular \emph{Kepler} \citep{Borucki2010}, have revolutionised the study of extrasolar planets, finding more than a thousand confirmed planets, revealing their occurrence as function of orbital distance and planet radius, and providing the first estimates of $\eta_{\rm{E}}$ -- the fraction of stars that harbour a rocky planet in their habitable zone \citep{Fressin2013, Dressing2015}.

Since both CoRoT and \emph{Kepler} surveyed relatively small fractions of the sky the planet host stars typically have faint apparent magnitudes. This is partly mitigated by K2, the reincarnation of the \emph{Kepler} mission after two of the four reaction wheels of the spacecraft failed \citep{Howell2014}, which is covering nineteen \emph{Kepler} fields up to early 2019. On the other hand the ground-based surveys have now monitored a large fraction of the sky, however their saturation limits usually prevent the discovery of systems at $m_V \gtrsim 8.5$, and until recently almost all known transiting systems brighter than this limit were first found through their radial velocity variations.

The brightest transiting exoplanets are scientifically extremely valuable. First of all, radial velocity follow-up to determine their masses is viable even for systems transiting early-type stars, where rapid stellar rotation and few spectral lines pose a challenge. This also applies to observations of the spectroscopic transit to determine a system's spin-orbit alignment \citep{CollierCameron2010}. Secondly, atmospheric characterisation at high spectral resolution can only be achieved on bright systems. For example, ground-based high-dispersion detections of transmission signals from carbon monoxide and water have only been achieved for the bright HD\,209458 and HD\,189733 hot Jupiter systems \citep{Snellen2010,Brogi2016}. Thirdly, the stellar parameters of the host can be determined to a high accuracy, allowing in-depth analyses.

Mainly for these reasons, NASA is planning to launch the Transiting Exoplanet Survey Satellite (TESS) by the end of 2018. This first  all-sky transit survey will identify planets ranging from Earth-sized to gas giants, for a wide range of stellar types and orbital distances \citep{Ricker2015}. Ground-based initiatives to probe the bright end of the transiting exoplanet population include the Kilo-degree Extremely Little Telescope (KELT) survey \citep{Pepper2007}, which consists of two  wide-field $26\degr\times26\degr$ telescopes, in the northern and southern hemisphere, which have been operational since 2005 and 2009, respectively. KELT has, among other bright objects, discovered the brightest hot Jupiter system to date, KELT-9, with an apparent magnitude of $m_V=7.56$ \citep{Gaudi2017}. 

This paper describes the discovery of MASCARA-2\,b, the second planet discovered by the Multi-Site All-Sky CAmeRA \citep{Talens2017a} which previously discovered MASCARA-1\,b \citep{Talens2017b}. MASCARA-2\,b is a hot Jupiter transiting a $m_V=7.58$ A2 star. Section \ref{sec:mascara} describes MASCARA and the discovery observations of the new planetary system. Section \ref{sec:observations} describes the follow-up observations, including transit confirmation with the NITES and IAC80 telescopes, and radial velocity and spectroscopic transit observations with the SONG telescope. Section \ref{sec:analysis} presents the subsequent analysis to derive the system parameters of MASCARA-2, which are presented and discussed in Section \ref{sec:results}.

\section{MASCARA \& Discovery Observations}
\label{sec:mascara}

\begin{figure}
  \centering
  \includegraphics[width=8.5cm]{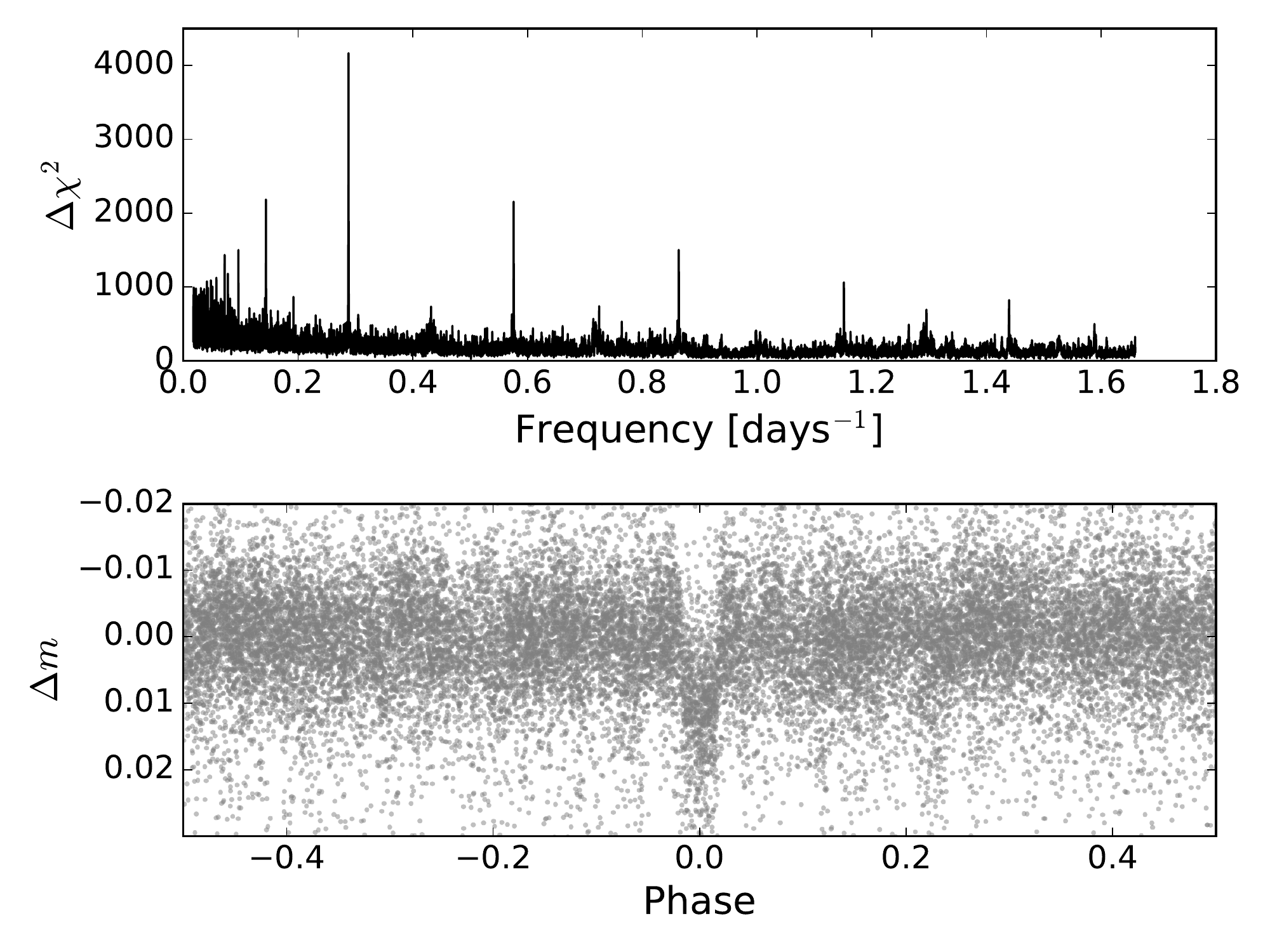}
  \caption{Discovery data of MASCARA-2\,b. \emph{Top panel}: BLS periodogram of the calibrated MASCARA light curve obtained between February 2015 and December 2016. The peak in the periodogram is located at a frequency of $0.288~\rm{days}^{-1}$. \emph{Bottom panel}: The calibrated MASCARA data, consisting of $24309$ points after binning to a cadence of $320~\rm{s}$. The data has been phase-folded to a period of $3.474~\rm{days}$, corresponding to the peak in the periodogram.}
  \label{fig:boxlstsq}
\end{figure}

The primary objective of the Multi-site All-Sky CAmeRA \citep[MASCARA;][]{Talens2017a} is to find transiting planetary systems at $m_V < 8$. It consists of two stations, one in the northern hemisphere at the Observatorio del Roque de los Muchachos, La Palma (Canary Islands, Spain), which has been operational since early 2015, and one in the southern hemisphere at the European Southern Observatory at La Silla (Chile), which saw first light in June 2017. Each station consists of five interline CCD cameras equipped with wide-field lenses that together monitor the entire local sky down to airmass 2 and partly down to airmass 3 \citep{Talens2017a}. The cameras point at fixed points on the local sky, meaning stars follow the same track across the CCDs each night. Images are taken with an exposure time of 6.4 seconds, short enough such that the motion of stars during an exposure is less than one pixel. Read-out of each image takes place during the next exposure so that no observing time is lost. The images are taken at the same sidereal times each night and astrometric solutions are updated every 50 images (${\sim}5$ minutes). The astrometric solutions are used to perform aperture photometry on all known stars brighter than $m_V < 8.4$. Due to the large data volume collected (${\sim}15$ terabytes per station per month) the raw data is overwritten after several weeks. A permanent record of the night sky is kept in the form of 50-frame averages of the raw images, allowing possible future searches for transients or other phenomena. Details of the subsequent analysis, converting the raw flux measurements into calibrated stellar light curves, will be described in detail in \citet[in prep.]{Talens2018}. In short, for each camera the position dependent transmission is determined from the combined light curves of all stars observed over a period of two weeks. Furthermore, the time dependent throughput of the Earth atmosphere (clouds, hazes, etc.) is determined for every image across a coarse sky-grid. In addition, each star exhibits a unique systematic trend as function of local siderial time, as a result of two effects. First, there are strong directional variations in the point-spread function (PSF) of the camera system, which cause variations in the contribution of possible neighbouring stars to the photometric aperture. Second, there are residual transmission variations due to differences in the colors of the stars. This PSF effect is corrected for each star individually by means of an empirical fit. The calibrated lightcurves are subsequently searched for transit events using the Box Least-Squares (BLS) algorithm \citep{Kovacs2002}.

MASCARA-2 (HD\,185603, HIP\,96618) has been observed by the northern MASCARA station since early 2015. Its nightly path along the sky passes across the East, Zenith, and West cameras. To date, more than 1.6 million photometric measurements have been taken, totalling almost 3000 hours of observing time. The final calibrated 5-minute binned light curve exhibits an (outlier rejected) dispersion of ${\sim}1.5\%$. The top panel of Fig. \ref{fig:boxlstsq} shows the BLS periodogram, revealing a strong peak at a frequency of $1/3.474~\rm{days}^{-1}$, the bottom panel shows the light curve, phase folded to this frequency. A transit-like signal is detected at $12.7\sigma$. 

\section{Follow-up Observations}
\label{sec:observations}

\begin{table*}
\centering
\caption{Observations used in the discovery of MASCARA-2\,b.}
\begin{tabular}{l c c c c c}
 Instrument & Observatory & Date & $N_{\rm{obs}}$ & $t_{\rm{exp}}$ [s] & Filter/Spectral range \\ 
 \hline
 \hline
 MASCARA & Roque de los Muchachos & Feb. 2015 - Dec. 2016 & 24309 & 320 & None \\  
 SONG & Teide & May-June 2017 & 16 & 600 & $4400-6900~\text{\AA}$\\
 SONG & Teide & 28 May 2017 & 19 & 600 & $4400-6900~\text{\AA}$\\
 SONG & Teide & 04 June 2017 & 30 & 600 & $4400-6900~\text{\AA}$\\
 NITES & Roque de los Muchachos & 04 June 2017 & 940 & 10 & R\\
 CAMELOT@IAC80 & Teide & 04 June 2017 & 398 & 40 & R\\
\end{tabular}
\label{tab:datasets}
\end{table*}

After the detection of the transit-like signal in the MASCARA data, several follow-up observations had to be taken to confirm the planetary nature of the system and to measure its main characteristics. First, follow-up photometric transit observations are required to reject the possibility that the transit-like signal is caused by variability in a background object in the few-arcminute vicinity of the bright star, and to refine the shape of the transit. Second, spectroscopic transit observations are needed to definitely establish that the transit occurs in front of the bright target star, and allow the determination of the projected spin-orbital alignment of the system. Finally, radial velocity measurements are required to determine the mass of the transiting object, or at least to demonstrate that it is within the planetary mass range and not a brown or red dwarf star. Table \ref{tab:datasets} lists all photometric and spectroscopic observations used in the discovery of MASCARA-2\,b. 

\subsection{Follow-up photometric observations}

\begin{table*}
\small
\centering
\caption{Parameters and best-fit values used in modelling the MASCARA, NITES and IAC80 photometric data.}
\begin{tabular}{l c c c c c c}
 Parameter & Symbol & Units & MASCARA & NITES & IAC80 & Joint\tablefootmark{a}\\
 \hline
 \hline
 Reduced chi-square & $\chi^2_\nu$ & - & 3.42 & 0.52 & 0.66 & 1.00 \\
 Epoch & $T_p$ & BJD & $2457909.593 \pm 0.001$ & $2457909.5905^{+0.0008}_{-0.0009}$ & $2457909.5907 \pm 0.0004$ & $2457909.5906^{+0.0003}_{-0.0002}$ \\  
 Period & $P$ & d & $3.474135 \pm 0.00001$ & $3.474135$ (fixed) & $3.474135$ (fixed) & $3.474119^{+0.000005}_{-0.000006}$ \\
 Duration\tablefootmark{b} & $T_{14}$ & h & $3.88^{+0.09}_{-0.07}$ & $3.51^{+0.10}_{-0.07}$ & $3.55 \pm 0.07$ & $3.55 \pm 0.03$ \\
 Planet-to-star ratio & $p=R_p/R_\star$ & - & $0.115 \pm 0.002$ & $0.115^{+0.006}_{-0.005}$ & $0.118^{+0.004}_{-0.003}$ & $0.115 \pm 0.001$ \\
 Impact parameter\tablefootmark{b} & $b$ & - & $0.64^{+0.05}_{-0.06}$ & $0.3 \pm 0.2$ & $0.46^{+0.07}_{-0.12}$ & $0.47^{+0.04}_{-0.06}$ \\ 
 Eccentricity & $e$ & - & 0 (fixed) & 0 (fixed) & 0 (fixed) & 0 (fixed) \\
\end{tabular}
\tablefoot{
\tablefoottext{a}{The errorbars were scaled so $\chi^2_\nu = 1$ for each individual dataset before performing the joint fit.}
\tablefoottext{b}{The best-fit values of $T_{14}$ and $b$ differ significantly between MASCARA and the Joint fit. This difference most likely originates from the algorithms used to calibrate the MASCARA data, which are known to potentially modify the transit shape.}
}
\label{tab:photpars}
\end{table*} 

On June 4, 2017 we observed a transit of MASCARA-2\,b with the 0.4m NITES telescope \citep{McCormac2014} on La Palma and the 0.8m IAC80 telescope\footnote{\url{http://www.iac.es/OOCC/instrumentation/iac80/}} at Observatorio del Teide. The NITES data consists of 940 images taken in Johnson-Bessel R-band filter, each with an exposure time of 10 seconds, with the telescope defocused to avoid saturation. The data are reduced in the same way as for the MASCARA-1\,b follow-up observations \citep{Talens2017b} using Python with CCDPROC \citep{Craig2015}, utilising a master bias, dark and flat. Non-variable comparison stars are then selected by hand and aperture photometry is extracted using SEP \citep{Barbary2016, Bertin1996}. The shift between each defocused image is measured using the DONUTS \citep{McCormac2013} algorithm, and the photometric apertures are re-centered between frames. The final aperture size is chosen to minimise the dispersion of the data points out of transit, resulting in an RMS of ${\sim}4.3~\rm{mmag}$.

The IAC80 data was obtained using the CAMELOT instrument \footnote{\url{http://www.iac.es/OOCC/wp-content/uploads/telescopes/IAC80/documentos/User_manual.pdf}}, and consists of 398 images also taken in the R filter. We used the standard 2-channel readout mode covering the full $10{\arcmin}\times10\arcmin$ field of view of the instrument. Each image had an exposure time of 40 seconds, and the telescope was extremely defocused to avoid saturation. The data were reduced using standard Pyraf routines performing aperture photometry after alignment of all the science frames. As in the case of NITES, the final aperture size is chosen to minimise the dispersion of the data points out of transit, resulting in an RMS in the light curve of ${\sim}1.6~\rm{mmag}$.

The light curves taken by IAC80 and NITES clearly show the transit event and the higher spatial resolution of these observations ($0.3-0.66$\arcsec/pixel versus $1$\arcmin/pixel for MASCARA) confirms that the transit is not caused by a non-related background eclipsing binary inside the MASCARA photometric aperture. 

\subsection{Follow-up spectroscopic observations}

Over the course of May and June 2017 sixteen spectra were obtained using the automated 1m Hertzsprung SONG telescope \citep{Andersen2014} at Observatorio del Teide, for the purpose of characterizing the host star and constraining the planet mass from the radial velocities. In addition, during the nights of May 28, 2017 and June 4, 2017 the SONG telescope observed one partial and one full transit, spanning a total of 3 and 5 hours respectively. For the second night of observations 1 hour around egress was missed since the star was too close to zenith to be observed with SONG. High-dispersion $R=90,000$ echelle spectra were obtained, covering the wavelength range from $4400$ to $6900~\text{\AA}$. All spectra were taken with 10 minute exposure times, utilising a slitwidth of 1.2 arcseconds. The spectra were extracted from the observations following the procedure outlined in \citet{Grundahl2017}, subsequently bad pixels were removed and the spectra were normalised. The reduced spectra have a typical signal-to-noise ratio of 75.

\section{System and Stellar Parameters}
\label{sec:analysis}

\subsection{Photometric transit fitting}
\label{ssec:phot}

\begin{figure}
  \centering
  \includegraphics[width=8.5cm]{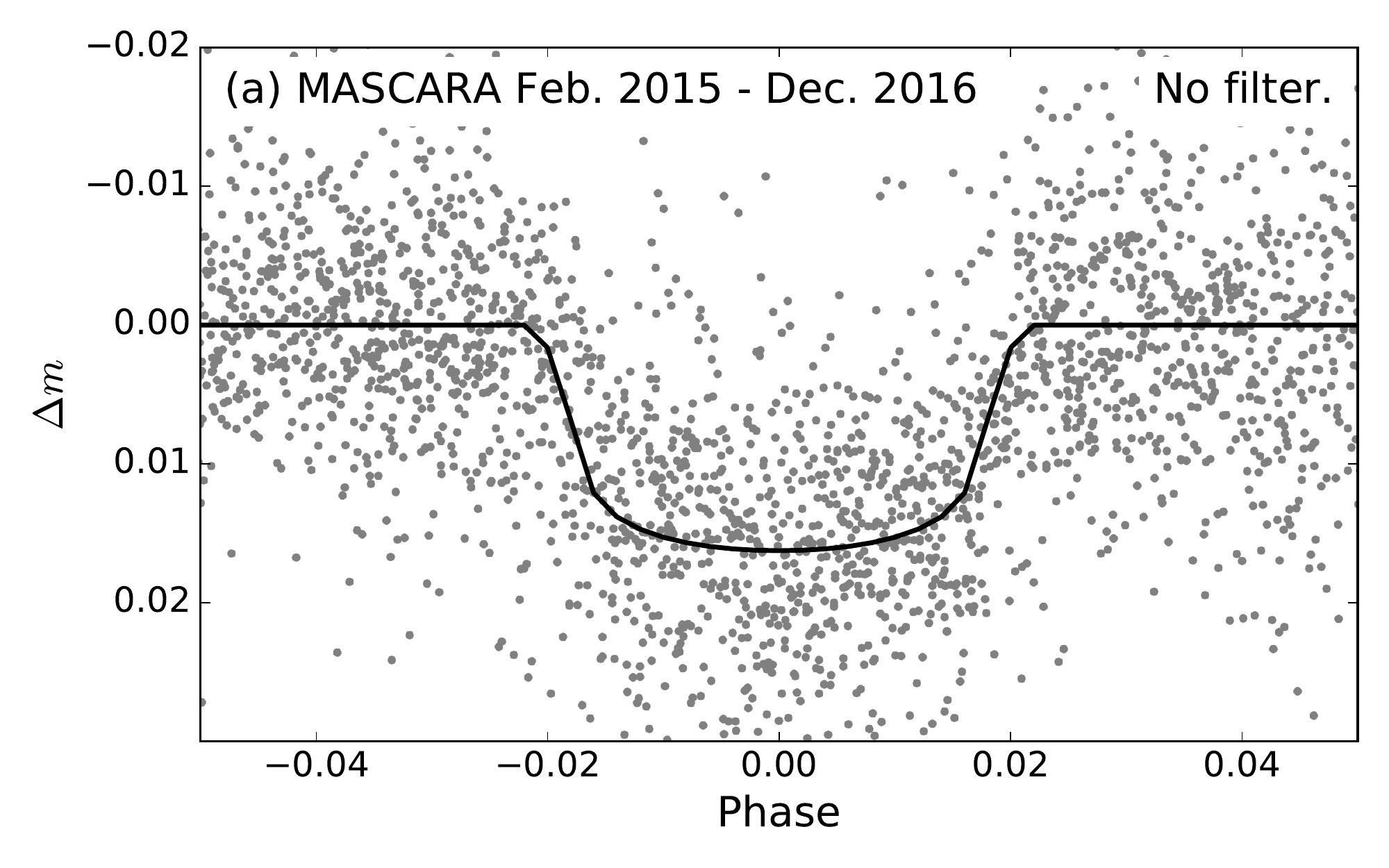}
  \includegraphics[width=8.5cm]{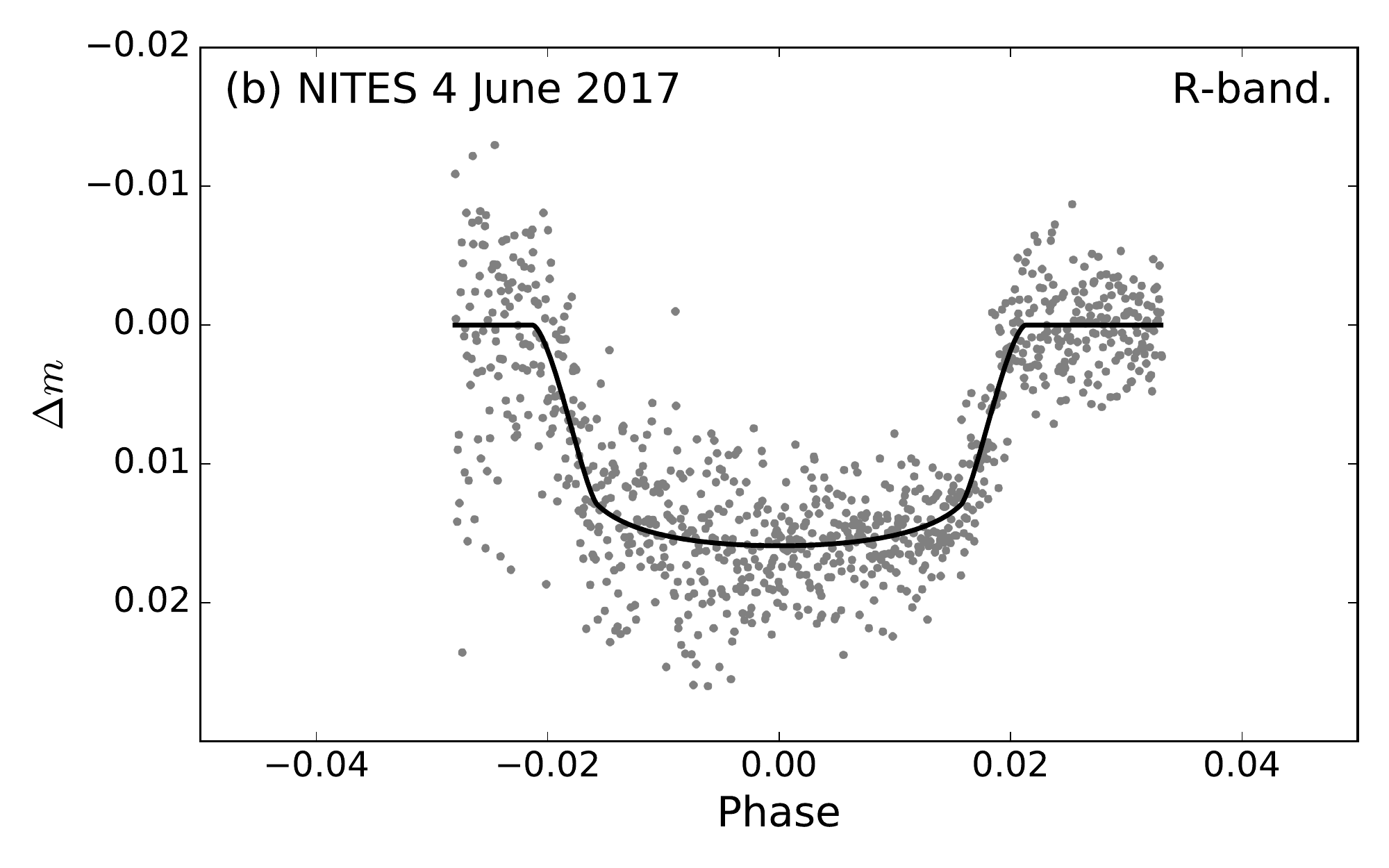}
  \includegraphics[width=8.5cm]{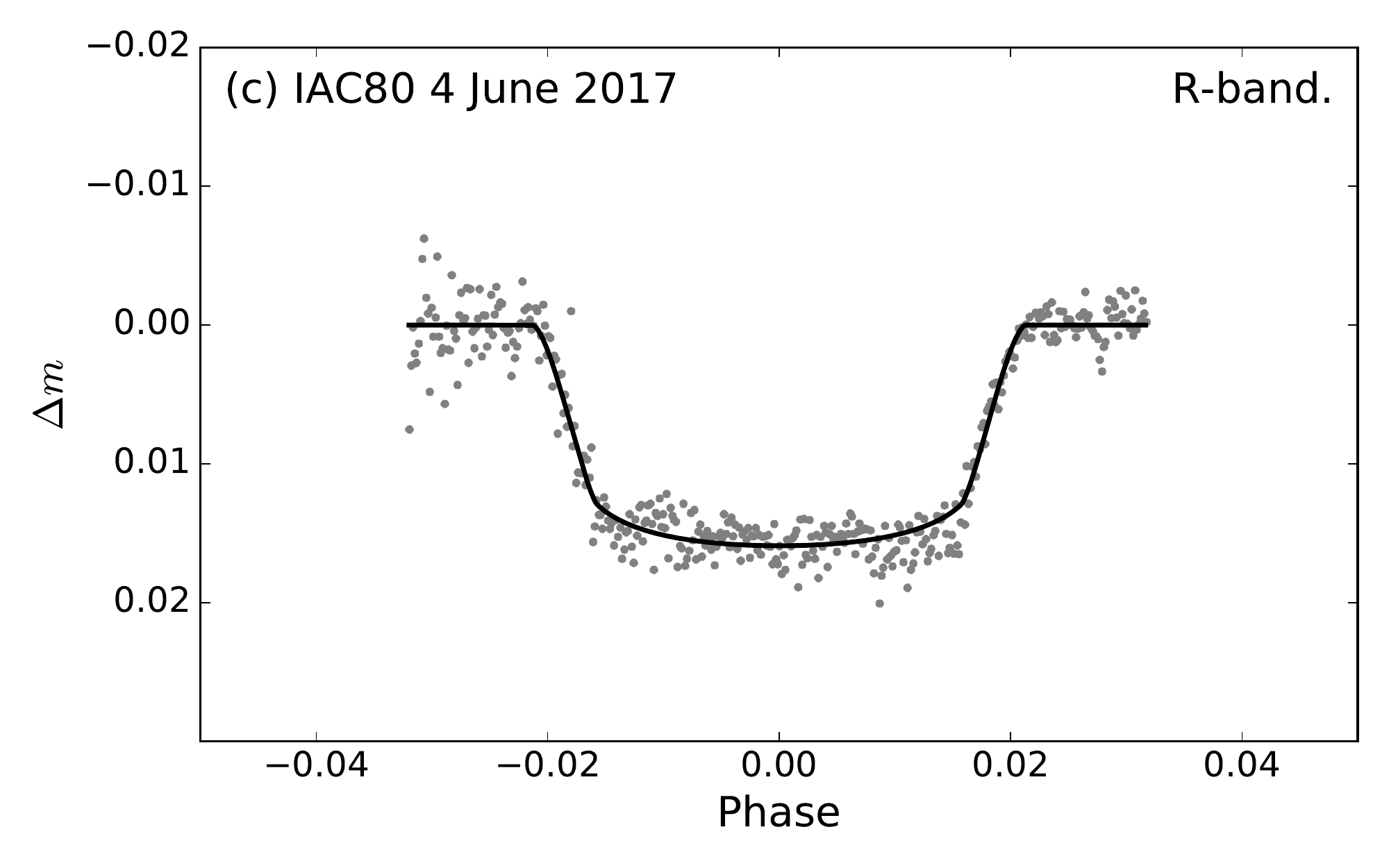}
  \caption{Photometric data (grey points) after subtraction of the best-fit baseline model, and best-fit transit model (black line) for MASCARA-2\,b. \emph{(a}) The calibrated MASCARA data obtained from February 2015 to December 2016. \emph{(b)} The NITES data obtained during the night of the 04 June 2017. \emph{(c)} The IAC80  data obtained during the night of the 04 June 2017.}
  \label{fig:photometry}
\end{figure}

\begin{figure}
  \centering
  \includegraphics[width=8.5cm]{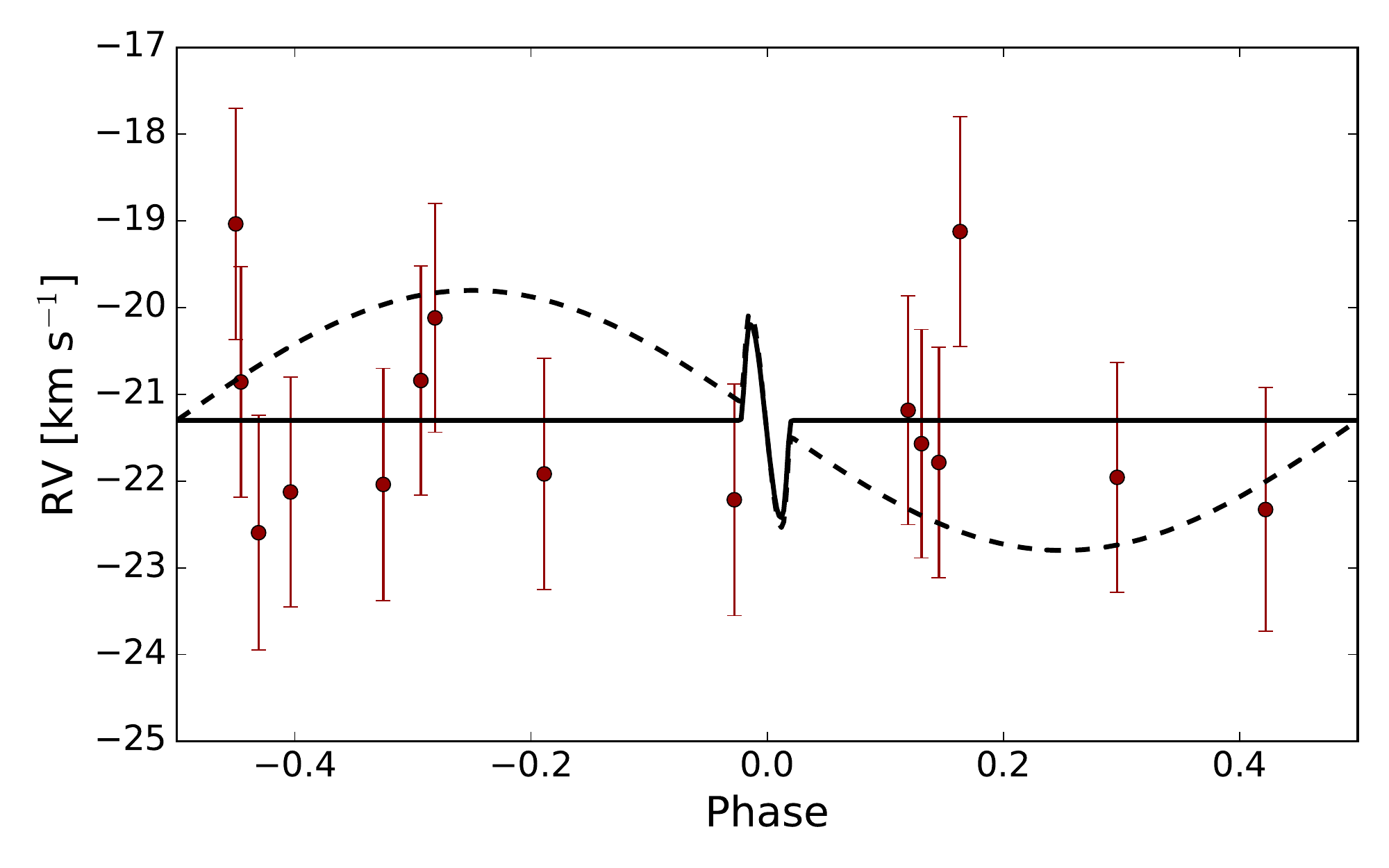}
  \caption{SONG RV data (red points) derived from the out-of-transit spectra, with the jitter term included in the uncertainties, best-fit model (solid black line) and $99\%$ upper limit (dashed black line) for the RV variations of MASCARA-2. The high rotational velocity of the host star, $\vsini~\rm{km~s}^{-1}$ broadens the spectral lines, reducing the accuracy of the individual RV points.}
  \label{fig:rv_points}
\end{figure}

The modelling of the photometric transit is performed in a similar way as for MASCARA-1\,b \citep{Talens2017b}. A \citet{Mandel2002} model is fitted to the MASCARA data using a Markov-chain Monte Carlo (MCMC) approach using the Python packages {\sc batman} and {\sc emcee} \citep{Kreidberg2015, FM2013}. We use a circular transit model ($e=0$), optimising for the transit epoch $T_p$, the orbital period $P$, the transit duration $T_{14}$, the planet-to-star radius ratio $p$ and the impact parameter $b$ using uniform priors on all parameters. For the baseline we choose to include the polynomials used to correct for the PSF effect (Sect. \ref{sec:mascara}) in our fitting routines, and fit the uncorrected MASCARA data. We employ a quadratic limb darkening law with fixed coefficients for V-band from \citet{Claret2011}, using values of $u_1 = 0.2752$ and $u_2 = 0.3179$, appropriate for the host star. Subsequently, the NITES and IAC80 photometric data sets are also fitted using the same model, except with the orbital period fixed to the best fit value obtained from the MASCARA photometry and a quadratic baseline model. Since the NITES and IAC80 data were taken in R-band, they were fitted with the R-band limb darkening coefficients from \citet{Claret2011} of $u_1 = 0.2163$ and $u_2 = 0.2814$. The best-fit parameter values and uncertainties were obtained from the median and the 16th and 84th percentiles of the output MCMC chains.

Table \ref{tab:photpars} lists the reduced chi-square and best-fit parameters for each dataset. The large reduced chi-square of the MASCARA data indicates that the errorbars are likely underestimated while those on the NITES and IAC80 data are overestimated. To obtain reliable values and uncertainties on the transit parameters we scaled the errorbars so $\chi^2_\nu = 1$ for each individual dataset before performing a joint fit to all photometric data. The resulting best-fit parameters are listed in Table \ref{tab:photpars} while Fig. \ref{fig:photometry} shows the photometry, after subtracting the baseline model, and best-fit transit model. The best-fit parameters of the joint fit are in good agreement with the fits to the individual datasets, with the exception of $T_{14}$ and $b$, which are significantly longer and higher for the MASCARA fit. We believe this is not surprising. Part of this tension can be explained by the underestimated errorbars producing underestimated uncertainties on the best-fit parameters of the MASCARA fit. Furthermore, the baseline model of the MASCARA data contains considerable freedom, and in the joint fit the NITES and IAC80 contain more information on the transit shape, driving the baseline solution. This freedom in the MASCARA shape stems from the significant data manipulations required in the calibration of the MASCARA data utilizing algorithms that are known to modify the shape of the transit. From the best-fit transit parameters we obtain a measurement of the stellar density, $\rho_\odot = 0.68^{+0.07}_{-0.06}~\rm{g~cm}^{-3}$

\begin{table*}
\centering
\caption{Parameters and best-fit values used in modelling the SONG RV and RM data.}
\begin{tabular}{l c c c c c c c}
 Parameter & Symbol & Units & RV & RM \\
 \hline
 \hline
 Epoch & $T_p$ & BJD  & $2457909.5906$ (fixed) & $2457058.42722 \pm 0.0003$ \\
 Period & $P$ & d & $3.474119$ (fixed) & 3.474135 (fixed) \\
 Scaled semi-major axis & $a/R_\star$ & - & - & $7.5 \pm 0.04$ \\
 Planet-to-star ratio & $p=R_p/R_\star$ & - & - & $0.1133 \pm 0.0007$ \\
 Cosine of the inclination & $\cos i$ & - & - & $0.066 \pm 0.002$ \\
 Eccentricity & $e$ & - & 0 (fixed) & 0 (fixed) \\
 RV amplitude\tablefootmark{a} & $K$ & km s$^{-1}$ & $0 \pm 600$ & $-1000 \pm 300$ \\
 Systemic velocity\tablefootmark{a} & $\gamma$  & km s$^{-1}$ & $-21.3 \pm 0.4$ & $-21.07 \pm 0.03$\\
 RV jitter & $\sigma_j$ & km s$^{-1}$ & $1.3^{+0.5}_{-0.3}$ & - \\
 Projected obliquity & $\lambda$ & \degr & - & $0.6 \pm 4$ \\
 Micro turbulence & $\nu$ & km s$^{-1}$ & - & $2.0 \pm 0.5$ \\
 Macro turbulence & $\zeta$ & km s$^{-1}$ & - & $15.1 \pm 0.6$ \\
 Projected rotation speed & $v \sin i_\star$ & km s$^{-1}$ & - & $114 \pm 3$ \\
 Limb darkening (SONG)\tablefootmark{b} & $u_1 + u_2$ & - & - & $0.63 \pm 0.01$ \\
 Limb darkening (IAC80)\tablefootmark{b} & $u_1 + u_2$ & - & - & $0.42 \pm 0.04$ \\ 
\end{tabular}
\tablefoot{
\tablefoottext{a}{Since the small phase coverage of the spectroscopic transit observations might cause short time-scale effects in the star, such as pulsations, to bias the observations we prefer the values for $K$ and $\gamma$ obtained from the RV modelling where such short time-scale effects are not coherent and included in the jitter term.}
\tablefoottext{b}{We employed Gaussian priors with a width of $0.05$ on $u_1 + u_2$ using the appropriate limb-darkening coefficients from \citet{Claret2011} for $u_1 + u_2$.}
}
\label{tab:specpars}
\end{table*}

\subsection{Stellar parameters determination}

A great benefit for the characterisation of transiting systems in MASCARA is that for most stars the basic properties are already available in the literature at these bright apparent magnitudes - as is the case for MASCARA-2 (HD\,185603; HIP\,96618). Hipparcos \citep{vanLeeuwen2007} lists a parallax of $8.73 \pm 0.5~\rm{mas}$, which is refined by the first data release of GAIA \citep{Lindegren2016} to $7.16 \pm 0.34~\rm{mas}$, corresponding to a distance of $140 \pm 7~\rm{pc}$. Just as for MASCARA-1, \citet{McDonald2012} determined the effective temperature and luminosity of MASCARA-2 (as for more than 100,000 other bright stars) by combining the photometry of a number of surveys  with the Hipparcos distance. They find an effective temperature of $T_{\rm{eff}} = 8809~\rm{K}$ and a bolometric luminosity (corrected to the GAIA parallax) of $15.1~\rm{L}_{\odot}$, but provide no uncertainties. 

From an analysis of the SONG spectra with iSpec \citep{BC2014} we independently determine the stellar parameters. For this analysis we first use BASTA \citep{SilvaAguirre2015} with a grid of BaSTI isochrones \citep{Pietrinferni2004} to convert the photometric stellar density to a prior on the surface gravity. We modify iSpec to include the $\log g$ prior in the spectroscopic fitting process and derive the effective temperature, surface gravity, metallicity and projected rotation speed of the host star. Using BASTA, we then fit the spectroscopically derived effective temperature and metallicity with the photometric stellar density to derive the final stellar parameters.

The resulting parameters are $T_{\rm{eff}} = \teff~\rm{K}$, $\log g = \logg$, $[\rm{Fe}/\rm{H}] = \metal$ and $v\sin i_\star = 117 \pm 3 ~\rm{km~s}^{-1}$. The SONG effective temperature lies within $1\sigma$ of the \citet{McDonald2012} value. Since the latter do not quote uncertainties, and taking into account the precision of the available GAIA parallax, we assume an effective temperature of $T_{\rm{eff}} = \teff~\rm{K}$ and a bolometric luminosity of $15 \pm 3 ~\rm{L}_{\odot}$ for the remainder of the analysis.

We determine the stellar mass, radius and age to be $\mstar~\rm{M}_\odot$, $\rstar~\rm{R}_\odot$ and $\age~\rm{Myr}$. From the bolometric luminosity and effective temperature, we independently derive a stellar radius of $R_{\star} = 1.7 \pm 0.2~\rm{R}_{\odot}$, consistent with the spectroscopically determined radius.

\subsection{Radial velocity analysis}

\begin{figure*}
  \centering
  \includegraphics[width=8.5cm]{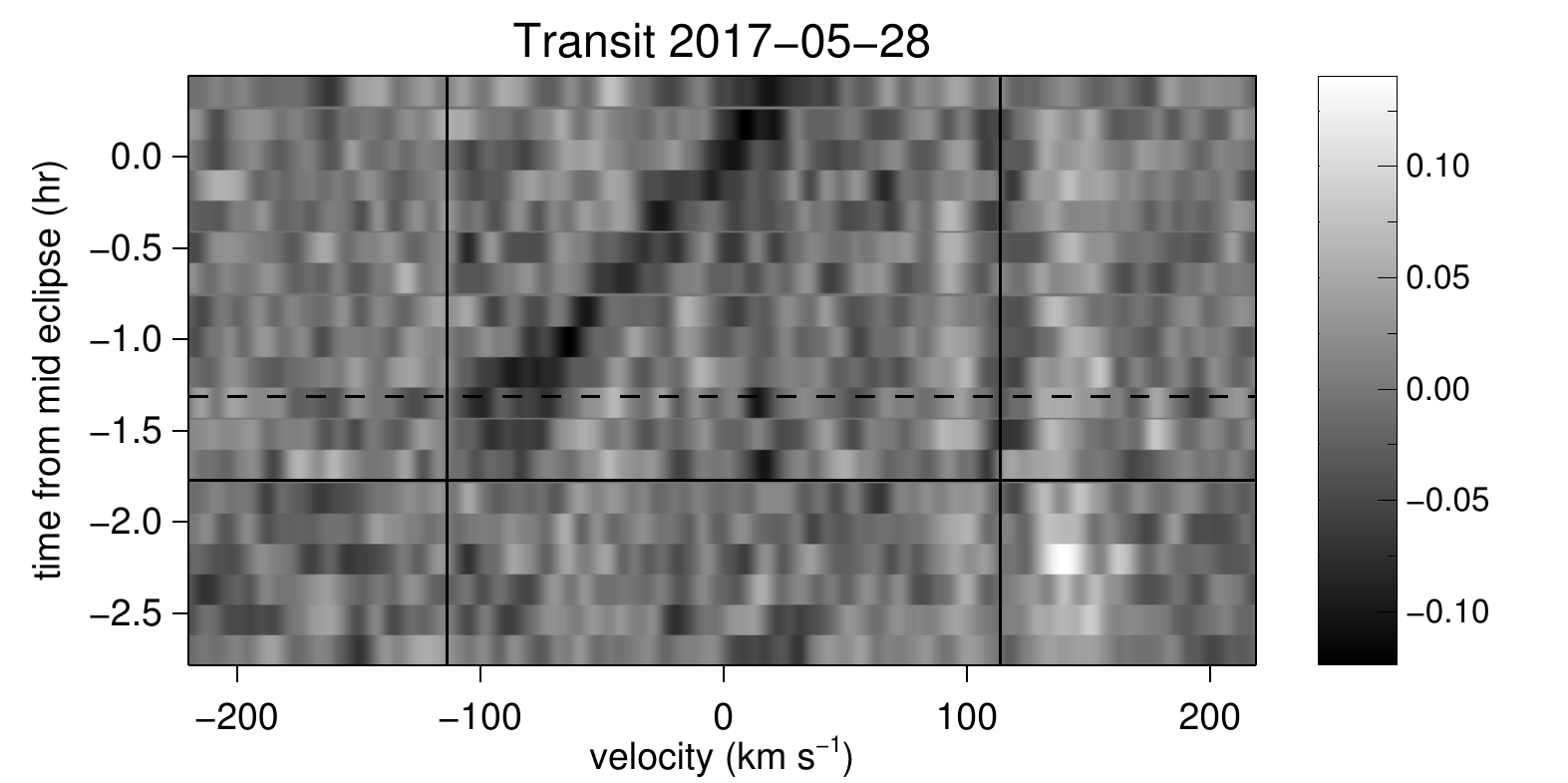}
  \includegraphics[width=8.5cm]{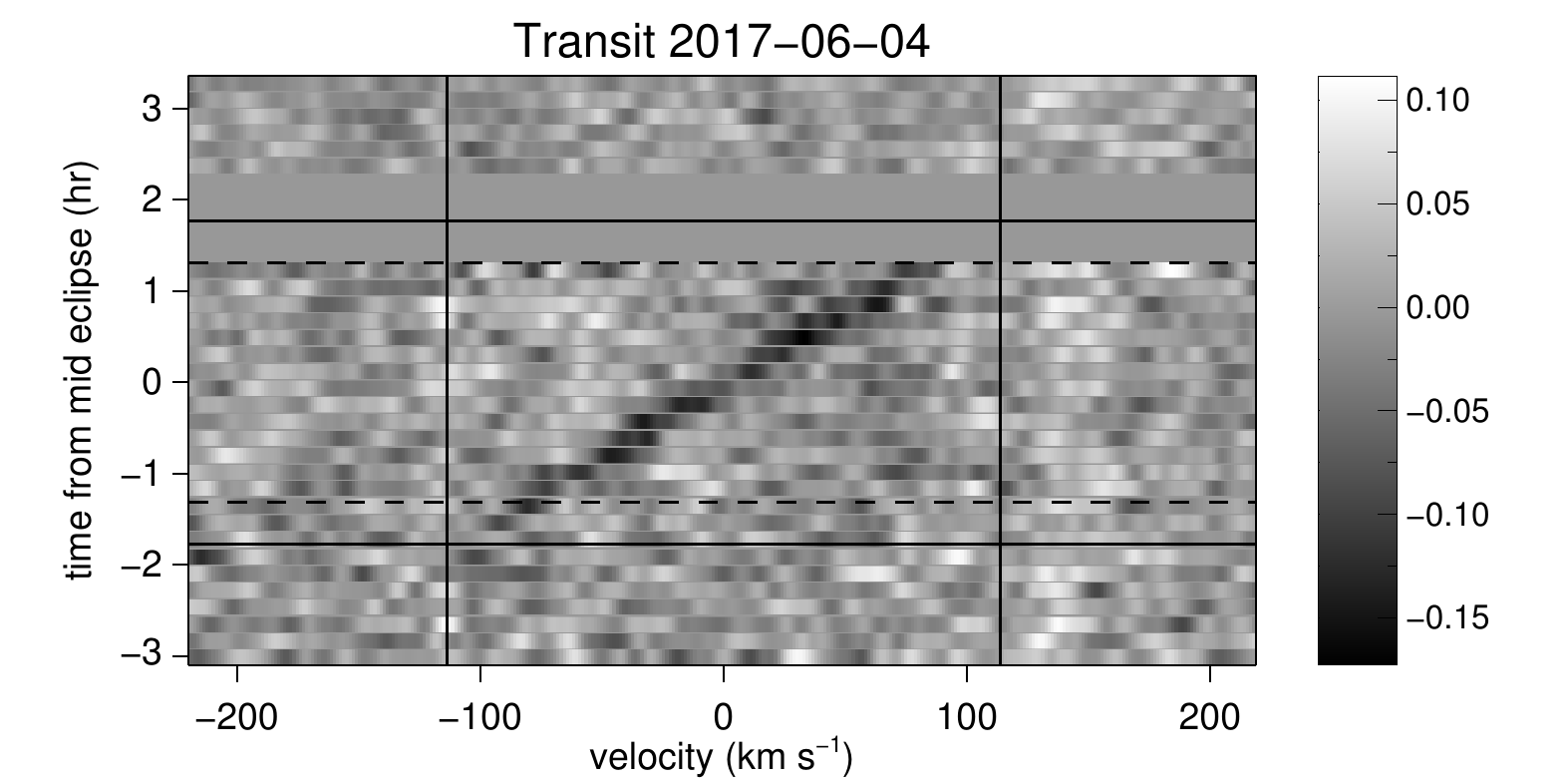}
  \includegraphics[width=8.5cm]{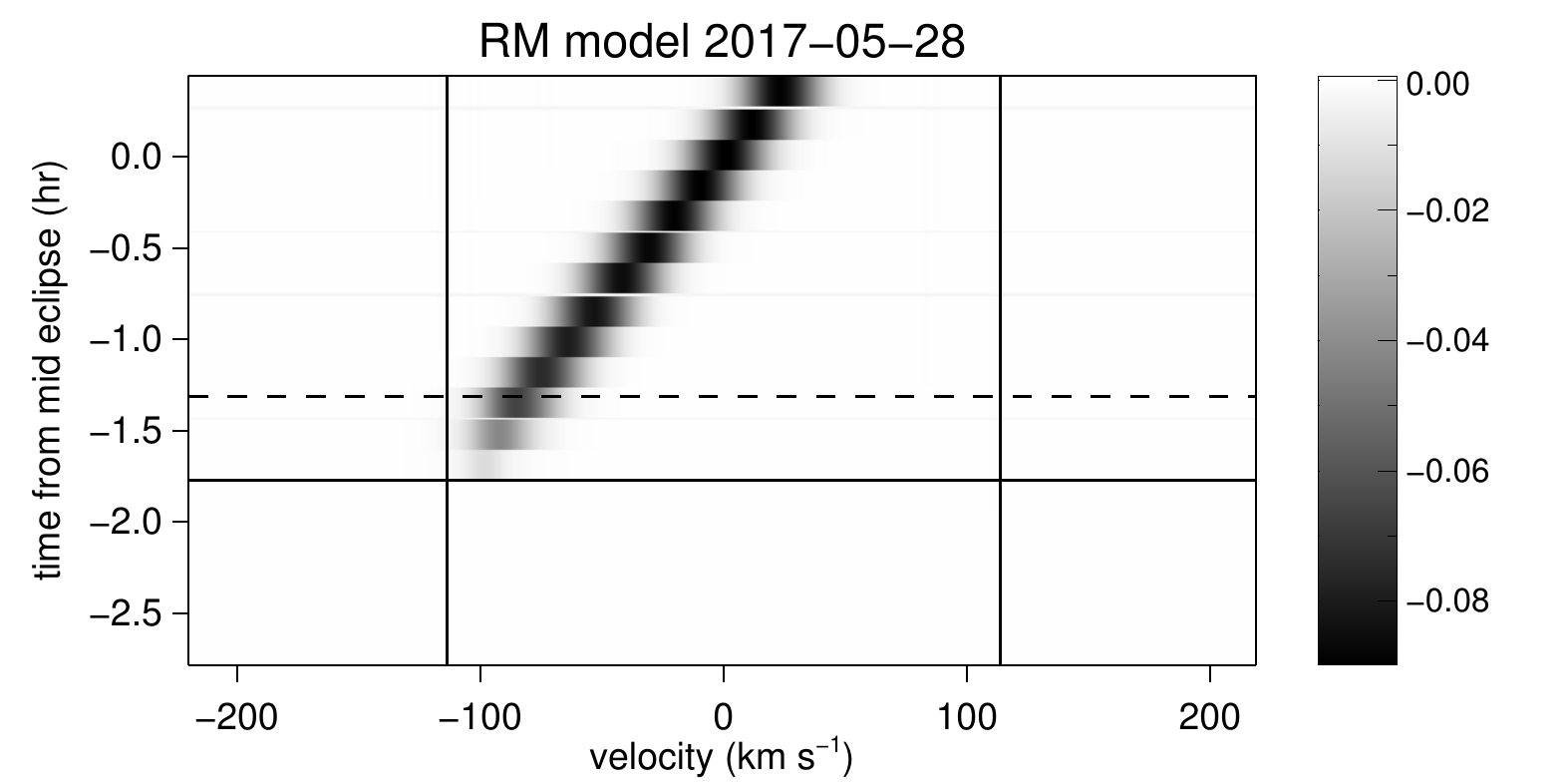}
  \includegraphics[width=8.5cm]{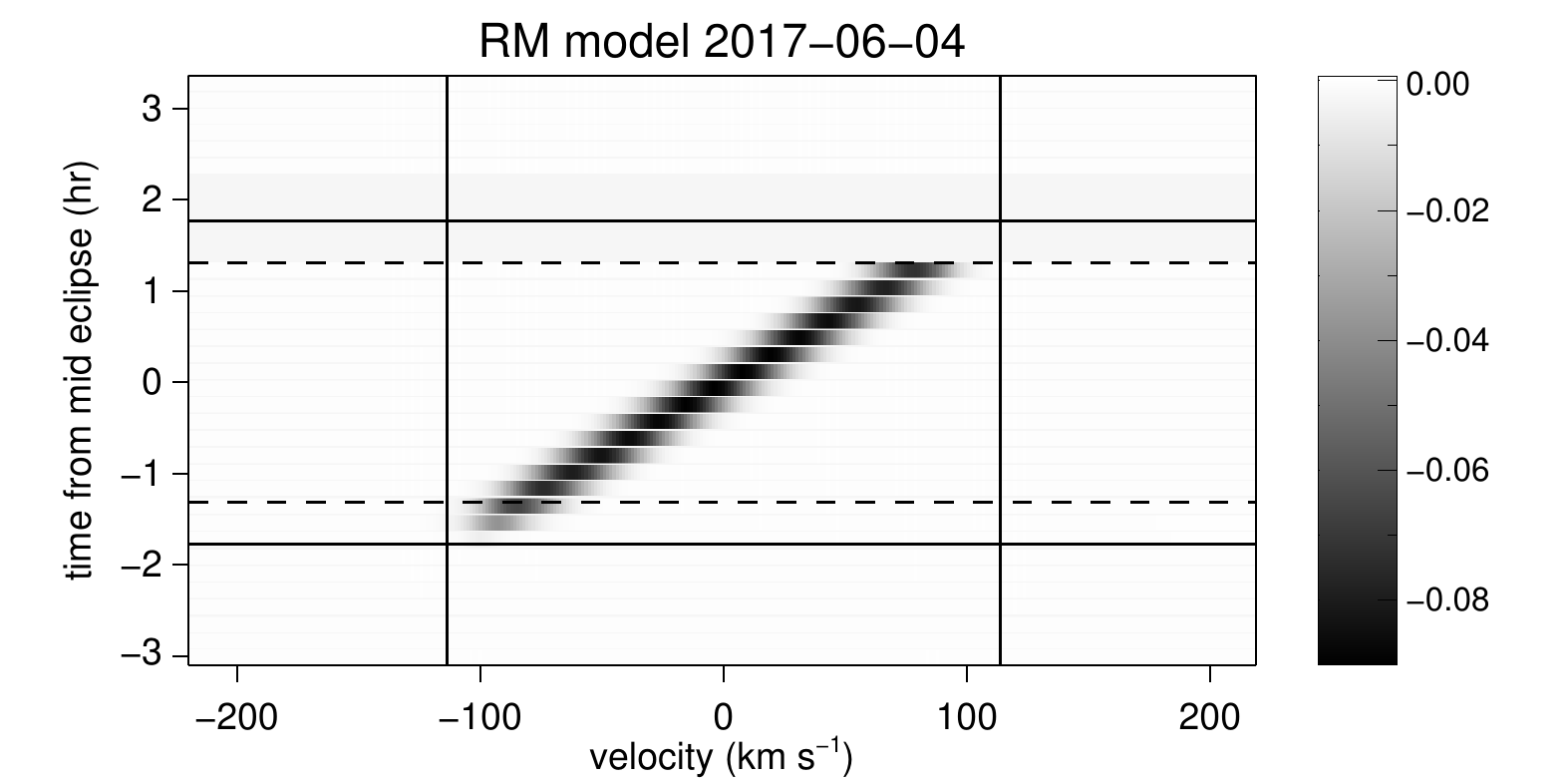}
  \includegraphics[width=8.5cm]{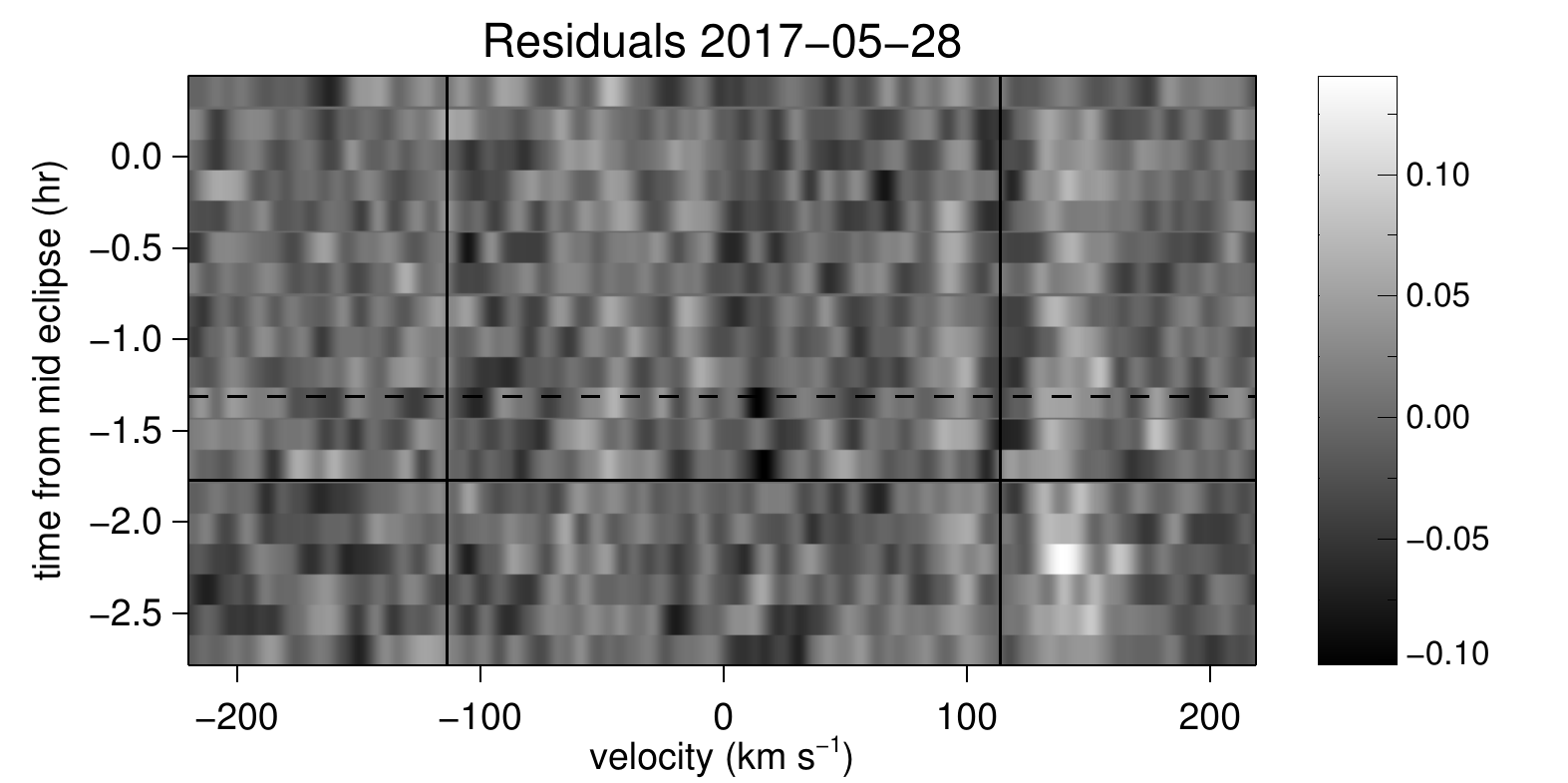}
  \includegraphics[width=8.5cm]{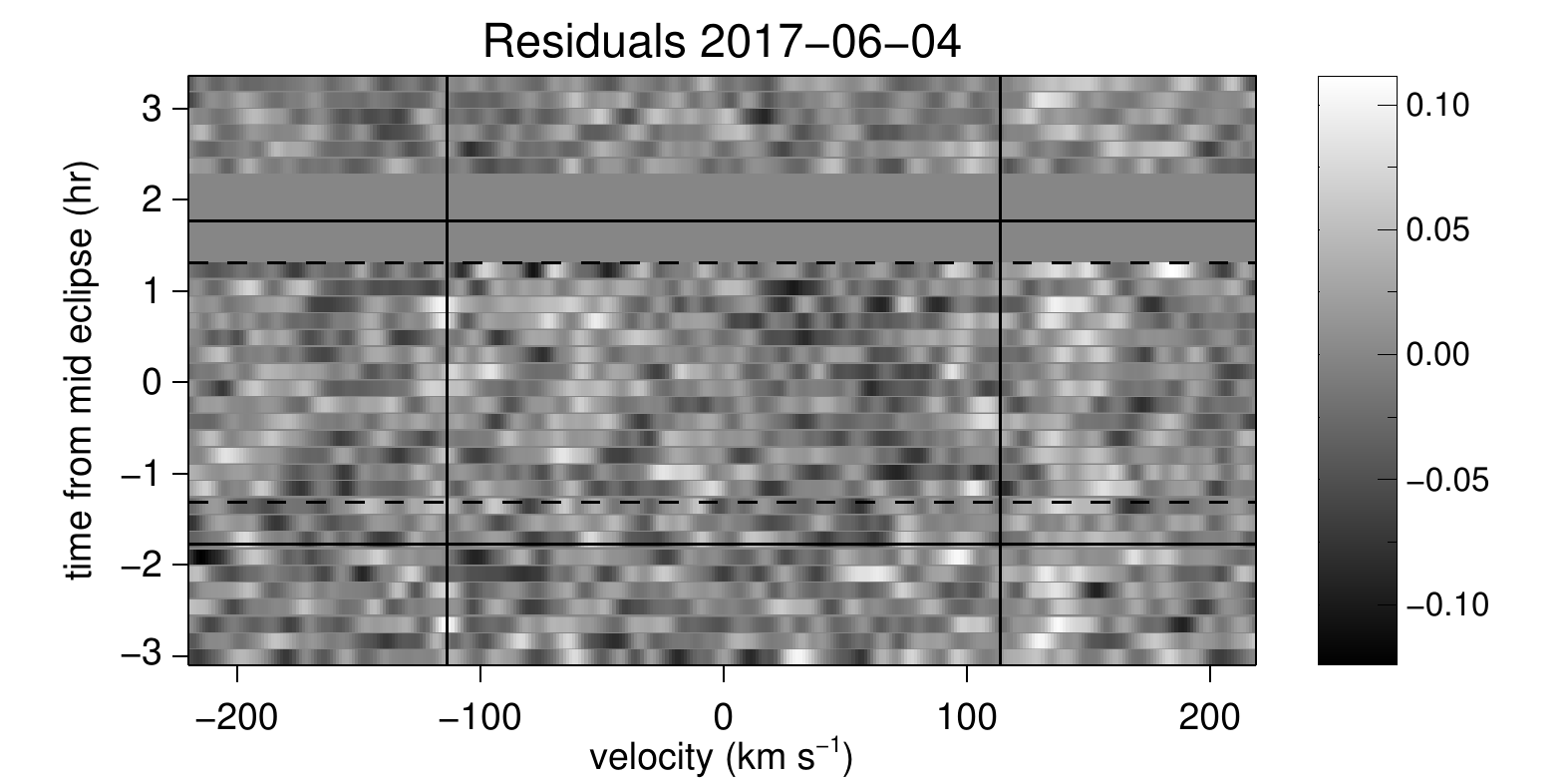}
  \caption{Rossiter-McLauglin measurements of MASCARA-2 during the transits occuring in the nights of May 28 (\emph{left column}) and June 4 (\emph{right column}). On June 4 one hour around egress was missed since the star was too close to zenith to be observed with SONG. In each panel the solid horizontal lines indicate the beginning of ingress and the end of egress and the dashed horizontal lines indicate the end of ingress and the beginning of egress. The vertical lines indicate the width of the best-fit rotational broadening profile. Shown in the rows are the observed CCFs after subtracting the average line profile and revealing the planet shadow (\emph{top}), the best-fit model to the observations (\emph{middle}) and the residuals of the observations after subtracting the best-fit model. (\emph{bottom}).}
  \label{fig:rm_effect}
\end{figure*}

For the RV analysis we use the spectra taken out-of-transit, 15 in total, excluding the spectra taken for the RM measurements and one in-transit spectrum taken at another time. The RM data are excluded because their small phase coverage might cause short time-scale effects in the star, such as pulsations, to bias the results while for the out-of-transit data such short time-scale effects are not coherent and can be accounted for using a jitter term. RVs were obtained from the spectra by cross-correlating the individual spectral orders with a stellar template corresponding to the best-fit spectroscopic parameters, generated using MOOG and assuming a Kurucz ATLAS9 stellar atmosphere \citep{Kirby2011}. The resulting cross-correlation functions (CCFs) were subsequently fitted using a rotationally broadened line profile to determine velocity shifts. The final RV measurement of each observation was determined from the weighted-mean of the measurements from the individual orders.

Since the projected equatorial rotation velocity of the star is very high rotational broadening is the dominant component in the line-profile, resulting in substantial uncertainties on the RV measurements of ${\sim}300~\rm{m~s}^{-1}$, which combined with a jitter term results in a typical uncertainty of ${\sim}1~\rm{km~s}^{-1}$. The RV data points are phase folded using the best-fit period and epoch as derived from the photometry, and a circular orbital solution is fitted to the data, keeping the period and epoch fixed to the photometric values and allowing the RV amplitude $K$, systemic velocity $\gamma$ and the jitter term $\sigma_j$ to vary. 

Table \ref{tab:specpars} lists the best-fit parameters and Fig. \ref{fig:rv_points} shows the RV data and the best-fit model, including the RM variations described in the next section. We are unable to constrain the amplitude of the RV variations, finding $K = 0.0 \pm 0.6~\rm{km~s}^{-1}$ (solid line in Fig. \ref{fig:rv_points}) and instead place a $99\%$ upper limit of $K < 1.5~\rm{km~s}^{-1}$ (dashed line in Fig. \ref{fig:rv_points}), corresponding to a limit on the companion mass of $M_p \mplan~\rm{M}_{\rm{J}}$, indicating that the companion is indeed in the planetary mass regime.

\subsection{Spectroscopic transit analysis}
\label{sec:rm_model}

A crucial element in confirming the planetary nature of the transiting system is the measurement of the spectroscopic transit, which confirms that a dark object is transiting the bright, fast spinning star, and in addition determines the projected spin-orbit angle of the system. Analysis of the SONG data was performed in a similar way as for the MASCARA-1 system \citep{Talens2017b}. CCFs were created from the SONG spectra using a stellar template as described in the previous section and the individual orders were co-added to obtain a single CCF for each observation. The resulting CCFs are shown in the top row of Fig. \ref{fig:rm_effect}, after subtraction of the average line profile.

We used the code from \citet{Albrecht2013} as described in \citet{Talens2017b} to measure the projected obliquity, $\lambda$, from the CCFs. As for the MASCARA-1 analysis, an MCMC was run to obtain confidence intervals for $\lambda$ and $v\sin i_\star$, while also optimizing the micro- and macro-turbulence, the transit epoch $T_p$, the planet-to-star radius ratio $p$, scaled semi-major axis $a/R_{\star}$, the cosine of the orbital inclination $\cos i$, the stellar systemic velocity $\gamma$, the stellar RV amplitude $K$, and the quadratic limb darkening parameters. For the spectroscopic observations we take limb darkening parameters of $u_1 = 0.27$ and $u_2 = 0.32$ \citep{Claret2011} and allow $u_1 + u_2$ to vary with a Gaussian prior of $0.59 \pm 0.05$. While running the MCMC on the 49 CCFs obtained during the two transit nights we simultaneously fit the IAC80 photometry. This adds the stellar limb darkening in the R band as an additional free parameter. Here we use the same R band limb darkening as in Sect. \ref{ssec:phot} to set a Gaussian prior of $u_1 + u_2 = 0.4977 \pm 0.05$ while keeping $u_1 - u_2$ fixed. Each time the likelihood is calculated, two free parameters are optimised for each observation allowing for offsets and scaling in intensity. This should include possible effects of non-perfect normalisation and the changing signal-to-noise of the spectra.

The middle and bottom rows of Fig. \ref{fig:rm_effect} show the best-fit models and their residuals. We detect a clear signal from the planet as it crosses the disk of the star and determine a projected spin-orbit angle of $0.6 \pm 0.5 \degr$ indicating that the system is aligned. The orbital parameters $T_p$, $a/R_\star$, $p$ and $\cos i$ are in good agreement with those derived from photometry alone, and provide us with an improved measurement of the stellar density $\rho_\odot = \sdens~\rm{g~cm}^{-3}$, consistent with the value derived from the photometry. The derived value of $v \sin i_\star = 113.6 \pm 0.2~\rm{km~s}^{-1}$ is somewhat lower than the spectroscopic value which we attribute to the inclusion of macro-turbulence here. We argue that the formal uncertainties in the projected obliquity and projected rotation speed underestimate the true uncertainty in these parameters. Since the host star is a rapid rotator we expect both a departure from spherical symmetry and the presence of gravity darkening, whose effects are not included in the modelling presented here. We therefore argue that uncertainties of $4 \degr$ and $3~\rm{km~s}^{-1}$ are more appropriate for the projected obliquity and projected rotation speed.

\section{Discussion and Conclusion}
\label{sec:results}

\begin{table}
\small
\centering
\caption{Parameters describing the MASCARA-2 system, derived from the best-fit models to the photometric and spectroscopic data.}
\begin{tabular}{l c c c}
 Parameter & Symbol & Value \\
 \hline
 \hline
 Stellar parameters \\
 \hline
 Identifiers & & HD\,185603; HIP\,96618 \\
 Right Ascension & & $19^h38^m38.73^s$ \\
 Declination & & $+31\degr13\arcmin09.2\arcsec$ \\
 Spectral Type & & A2\\
 V-band magnitude & $m_V$ & 7.6 \\
 Age & & $\age~\rm{Myr}$ \\
 Effective temperature & $T_{\rm{eff}}$ & $\teff~\rm{K}$ \\
 Projected rotation speed & $v\sin i_\star$ & $\vsini~\rm{km~s}^{-1}$ \\
 Surface gravity & $\log g$ & \logg \\
 Metallicity & [Fe/H] & \metal \\  
 Stellar mass & $M_\star$ & $\mstar~\rm{M}_{\sun}$ \\
 Stellar radius & $R_\star$ & $\rstar~\rm{R}_{\sun}$ \\
 Stellar density & $\rho_\star$ & $\sdens~\rm{g~cm}^{-3}$ \\
 \hline
 Planet parameters\\
 \hline
 Planet radius & $R_p$ & $\rplan~\rm{R}_{\rm{J}}$ \\
 Planet mass & $M_p$ & $\mplan~\rm{M}_{\rm{J}}$ \\
 Equilibrium temperature\tablefootmark{a} & $T_{\rm{eq}}$ & $\teq~\rm{K}$ \\
 \hline
 System parameters\\
 \hline  
 Epoch & $T_p$ & $\epoch~\rm{BJD}$ \\
 Period & $P$ & $\period~\rm{days}$ \\
 Semi-major axis & $a$ & $\axis~\rm{AU}$ \\
 Inclination & $i$ & $\inc~\degr$ \\
 Eccentricity & $e$ & \ecc \\
 Projected obliquity & $\lambda$ & $\obl~\degr$ \\
 \hline
\end{tabular}
\tablefoot{
\tablefoottext{a}{Computed assuming uniform redistribution and a Bond albedo of zero.}
}
\label{tab:syspars}
\end{table} 

We list the final parameters describing MASCARA-2\,b and its host star in Table \ref{tab:syspars}. We find that MASCARA-2\,b orbits its host star in $\period~\rm{days}$ at a distance of $\axis~\rm{AU}$. It has a radius of $\rplan~\rm{R}_{J}$ and we report a $99\%$ upper limit of $\mplan~\rm{M}_{J}$ for the mass. We continue to take RV measurements to further constrain the mass of the planet. 

MASCARA-2\,b orbits a hot ($T_{\rm{eff}} > 7000~\rm{K}$), early-type star, one of only eleven Hot Jupiters transiting hot stars known to date and of just four \citep[WASP-33\,b, KELT-9\,b, MASCARA-1\,b, MASCARA-2\,b,][]{CollierCameron2010,Gaudi2017,Talens2017b} orbiting stars brighter than $m_V = 8.4$. Of the fainter systems, three were discovered in the \emph{Kepler} field \citep[Kepler-13A\,b,Kepler-1171\,b,Kepler-1571\,b,][]{Mazeh2012,Morton2016}, three by KELT \citep[KELT-13\,b,KELT-17\,b,KELT-19A\,b,][]{Temple2017,Zhou2016,Siverd2017} and HATP-57\,b \citep{Hartman2015}. Less than halve of these planets have secure mass measurements and in general the sample is still too small to make a meaningful comparison of planet and orbital parameters to those of systems with late-type host stars. Surprisingly, the projected spin-orbit angle of MASCARA-2\,b is found to be consistent with zero, contrary to what is typically found for planets around early-type stars \citep{Winn2010,Schlaufman2010,Albrecht2012}. While we only measure the projected alignment, it is highly unlikely for a misaligned system to produce a projected obliquity of $\obl\degr$. It will be interesting to investigate why this system is aligned when similar systems are not.

Planet formation theory predicts that gas giant planets occur more frequently around more massive stars \citep{Ida2005,Kennedy2008}, since the mass of the planet forming disk scales with the mass of the host stars. Radial velocity surveys have confirmed this prediction for planets orbiting giant stars \citep[e.g.][]{Reffert2015}. However, gas giants planets are expected to form on wide orbits \citep{Lin1996,Bodenheimer2000}, requiring migration to shorter orbits for transit surveys to be able to detect them. It is not yet clear how stellar mass may influence the outcome of migration processes. In this context the observed obliquity distribution of transiting planets around massive early-type stars might provide constraints on which migration mechanism is responsible \citep{Fabrycky2007,Nagasawa2008}.

The brightness of its host star and its high equilibrium temperature make MASCARA-2\,b highly suitable for atmospheric studies. Using high-resolution transmission spectroscopy obtained from the ground we will be able to detect Na and K in the optical and CO and H$_2$O in the infrared. In particular the stellar alkali lines are significantly weaker than for solar-type stars, making the disentanglement of the planetary and stellar lines significantly easier. From space-based transit and eclipse spectroscopy with the HST and the JWST NIRISS instrument it will be possible to detect molecules such as CO, CH$_4$, NH$_3$ and H$^+$ and constrain the temperature-pressure profile of the atmosphere. However, the host star is of such brightness that it may staturate in other JWST instruments. Of particular interest is the possible detection of TiO, which is expected to be present in the atmospheres of highly irradiated ($T_{\rm{eq}} > 2000~\rm{K}$) planets and a possible cause of inversion layers. \citet{Nugroho2017} and \citet{Sedaghati2017} recently claimed detections of TiO in the atmospheres of WASP-33\,b and WASP-19\,b, using data obtained from the ground and space respectively. MASCARA-2\,b has an equilibrium temperature comparable to those of WASP-33\,b and WASP-19\,b making it an excellent target for attempting similar measurements.

The NASA TESS satellite \citep{Ricker2015} will also contribute to the characterization of the atmosphere, in particular it will measure the combination of dayside thermal and reflected starlight through the secondary eclipse, and their longitudinal distribution through the phase curve, constraining global circulation on this world.

We expect MASCARA to reveal many more bright transiting systems. Of the 14 transiting systems with $m_V<8.4$ known to date only two are located in the southern hemisphere. A second MASCARA station at the European Southern Observatory at La Silla, Chile, will start operations from July 2017 to unveil this planet population in the southern hemisphere.

\emph{During the preparation of this paper, our team became aware of another paper reporting the discovery of a planet orbiting HD\,185603 \citep{Lund2017}, published by the KELT collaboration. No information about any results or analysis was shared between the groups prior to the submission of both papers, and we would like the thank the KELT team for their collegiality and willingness to work with the MASCARA team to submit these discoveries simultaneously.}

\begin{acknowledgements}
IS acknowledges support from a NWO VICI grant (639.043.107). This project has received funding from the European Research Council (ERC) under the European Union's Horizon 2020 research and innovation programme (grant agreement nr. 694513). Simon Albrecht and Anders Justesen acknowledge support by the Danish Council for Independent Research, through a DFF Sapere Aude Starting Grant nr. 4181-00487B. Based on observations made with the Hertzsprung SONG telescope operated on the island of Tenerife by the Aarhus and Copenhagen Universities in the Spanish Observatorio del Teide of the Instituto de Astrofísica de Canarias. The Hertzsprung SONG telescope is funded by the Danish National Research Foundation, Villum Foundation, and Carlsberg Foundation. This research has made use of the SIMBAD database, operated at CDS, Strasbourg, France. This research has made use of the VizieR catalogue access tool, CDS, Strasbourg, France. We have benefited greatly from the publicly available programming language {\sc Python}, including the {\sc numpy, matplotlib, pyfits, scipy} and {\sc h5py} packages.
\end{acknowledgements}

\bibliographystyle{aa}
\bibliography{../mascara.bib}

\end{document}